\documentclass[aps,prl,twocolumn,floats,showpacs,superscriptaddress]{revtex4-1}
\usepackage{graphicx,epsfig}% Include figure files
\usepackage{times}
\usepackage{graphics,dcolumn,bm,float}
\usepackage{amssymb,amsmath,rotate,color}

\usepackage{graphicx}

\begin{document}
\unitlength 1 cm
\newcommand{\be}{\begin{equation}}
\newcommand{\ee}{\end{equation}}
\newcommand{\bearr}{\begin{eqnarray}}
\newcommand{\eearr}{\end{eqnarray}}
\newcommand{\nn}{\nonumber}
\newcommand{\la}{\langle}
\newcommand{\ra}{\rangle}
\newcommand{\cd}{c^\dagger}
\newcommand{\vd}{v^\dagger}
\newcommand{\ad}{a^\dagger}
\newcommand{\bd}{b^\dagger}
\newcommand{\tk}{{\tilde{k}}}
\newcommand{\tp}{{\tilde{p}}}
\newcommand{\tq}{{\tilde{q}}}
\newcommand{\eps}{\varepsilon}
\newcommand{\vk}{\vec k}
\newcommand{\vp}{\vec p}
\newcommand{\vq}{\vec q}
\newcommand{\vkp}{\vec {k'}}
\newcommand{\vpp}{\vec {p'}}
\newcommand{\vqp}{\vec {q'}}
\newcommand{\bk}{{\bf k}}
\newcommand{\bp}{{\bf p}}
\newcommand{\bq}{{\bf q}}
\newcommand{\br}{{\bf r}}
\newcommand{\bR}{{\bf R}}
\newcommand{\up}{\uparrow}
\newcommand{\down}{\downarrow}
\newcommand{\fns}{\footnotesize}
\newcommand{\ns}{\normalsize}
\newcommand{\cdag}{c^{\dagger}}

\title{Collective excitations and the nature of Mott transition in undoped gapped graphene}
\author{S. A. Jafari}
\affiliation{Department of Physics, Sharif University of Technology, Tehran 11155-9161, Iran}
\affiliation{School of Physics, Institute for Research in Fundamental Sciences (IPM), Tehran 19395-5531, Iran}
%\author{G. Baskaran}
%\affiliation{Institute of Mathematical Sciences, Chennai 600113, India}

\begin{abstract}
Particle-hole continuum (PHC) for massive Dirac fermions in presence of short range
interactions, provides an unprecedented
opportunity for formation of two collective split-off states, one in the singlet and 
the other in the triplet (spin-1) channel in undoped system. Both poles
are close in energy and are separated from the continuum 
of free particle-hole excitations by an energy scale of the
order of gap parameter $\Delta$. They both disperse linearly with
two different velocities reminiscent of spin-charge separation in
Luttinger liquids. When the strength of Hubbard 
interactions is stronger than a critical value, the velocity of
singlet excitation which we interpret as a charge boson composite becomes zero,
and renders the system a Mott insulator. Beyond this critical point,
the low-energy sector is left with a linearly dispersing 
triplet mode -- a characteristic of a Mott insulator. The velocity of triplet
mode at the Mott criticality is twice the velocity of underlying Dirac fermions.
The phase transition line in the space of $U$ and $\Delta$ is
in qualitative agreement with a more involved dynamical mean field theory (DMFT)
calculation.
\end{abstract}

\maketitle

\section{Introduction}
Graphene is a huge network of $sp^2$ bonds formed between carbon
atoms on a two-dimensional (2D) sheet~\cite{Novoselov}. 
The low-energy effective model for $p_z$
electrons in this system is the 2+1 dimensional Dirac theory which 
contains an energy scale proportional to $v_F$ -- the Fermi velocity of electrons~\cite{NetoRMP}. 
Effects of interactions and/or disorder can be taken into account 
on top of this non-interacting fixed point~\cite{NetoRMP}. Breaking the sub-lattice symmetry 
gives rise to a gap in the single-particle excitations spectrum~\cite{Lanzara}, 
which introduces the gap parameter $\Delta$ as another energy scale.  
The gapped graphene can be modeled with 2+1 dimensional massive Dirac
fermions~\cite{Alireza}. Gapped graphene supports interesting new class of excitations in
the spectrum such as, domain walls~\cite{Semenoff}. Moreover the single-particle
gap (mass) gives rise to suppression of Coulomb fields at nano-meter length 
scales~\cite{Pereira}. The gap could also give rise to universal nonlinear optical 
response~\cite{nloJafari}. When the many-body interactions of various form
are added to the massive Dirac theory, the nature of many-body excitations
becomes even more interesting. 
Recent {\em ab-initio} estimates of the strength of the 
Coulomb repulsion in various forms of graphene suggests that the
Hubbard parameter in graphene can be quite remarkable~\cite{Wehling}, which 
introduces a third energy scale, $U$. Therefore,  
an interesting question here would be the effects of local Hubbard type
interactions on the excitation spectrum of gapped graphene, and
the nature of transition to Mott insulating phase in gapped graphene.
Our recent full-fledged DMFT investigation of
the so called ionic-Hubbard model on the honeycomb lattice
suggested three phases~\cite{moradDMFT}: (i) Band insulating phase for $\Delta\gg U$.
(ii) Mott insulating phase for $\Delta\ll U$, and (iii) a semi-metallic
phase for $U\sim \Delta$. The above study is about the
nature of ground state. In order to study 
the connection of excited states and the ground state, 
one notes that; deep in the Mott insulating phase, 
the energy scale $U$ is expected to be dominant over $\Delta$, such that 
the low-energy excitations are expected to be spin fluctuations 
arising from super-exchange interaction induced by the large $U$. 
We would like to focus on, the evolution of the collective spin
and charge dynamics in this system as a function of the Hubbard $U$
for a system with a non-zero gap parameter $\Delta$. It turns out
that the presence of a non-zero gap parameter facilitates the
separation of two collective states each of which has a 
different velocity, reminiscent of the spin-charge separation in Luttinger liquids.

\section{Formulation of the problem}
The Hamiltonian we consider here is the so called ionic-Hubbard model,
which is defined by
\be
   H= \hbar v_F\sum_{\bk s} \psi^\dagger_{\bk s}\left[{\bf \sigma}.\bk+\sigma_z\Delta\right] \psi_{\bk s}
   +U\sum_{j} (n^a_{j\up}n^a_{j\down}+n^b_{j \up}n^b_{j\down})
\ee
where $n^f_{js}=f^\dagger_{js} f_{js}$ with $f=a,b$ corresponding to
number operator at the $j$'th unit cell at sub-lattices, A, B, respectively.
The spinor notation $\psi^\dagger_{\bk s}=(a^\dagger_{\bk s}~b^\dagger_{\bk s})$
has been used. Here $s=\up,\down$ denotes the $z$ component of the physical spin, 
$\bk=(k_x,k_y)$ is a two-dimensional momentum vector, and $\bf \sigma$
stands for Pauli matrices. $U$ is the strength of the on-site Coulomb
interaction known as Hubbard parameter, and $\Delta$ is the mass 
parameter. The quadratic part of this Hamiltonian can 
be diagonalized by a simple unitary transformation to the
basis of conduction ($+$) and valence ($-$) states, and the
corresponding eigen-values are given by,
\be
   \varepsilon^{\pm}(\bk)=\pm v_F\sqrt{|\hbar\bk|^2+\Delta^2},~~~~~~\Delta\ne 0,
   \label{massive.eqn}
\ee
where $\hbar v_F=\sqrt 3 t a/2$ and $a$ being the C-C bond length.
The hopping amplitude between the nearest neighboring carbon atoms
is $t\sim 2.8$ eV.
Due to paramagnetic nature of the non-interacting system (i.e. $U=0$), the 
occupation numbers and energies of the conduction and valence
bands are independent of spin orientation. We consider the undoped
system with precisely one $p_z$ electron per carbon atom, 
and assume the temperature to be zero. Then we extend the
collective mode analysis of Ref.~\cite{JafariBaskaranJPCM}
to the case with $\Delta\ne 0$, whereby the single-particle
spectrum changes from $\pm \hbar v_F|\bk|$ to the one given 
by Eq.~\eqref{massive.eqn}. 
This change in the single-particle spectrum, changes the borders
of the particle-hole continuum. However, as will be show here,
at the border corresponding to massless spectrum, interesting
collective quanta can be made in presence of short range Coulomb interactions. 
%figure 1
\begin{figure}[tb]
\begin{center}
\includegraphics[width=8cm,angle=0]{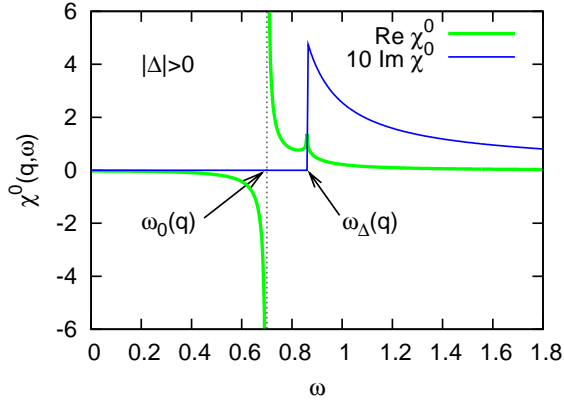}
\caption{ (Color online) Behavior of free particle hole 
propagator $\chi^0(\bq,\omega)$. The imaginary part has been magnified 
to clearly mark the onset of particle-hole continuum of free particle-hole 
excitations. In this figure, vertical scales are drawn 
for $qa=0.7$, and $v_F$ and $\hbar$ are assumed to be unit.
}
\label{chi0.fig}
\end{center}
\end{figure}

Using equation of motion it can be shown that a simple RPA-like
expression governs the eigen-value equation for the collective
excitations in the triplet and singlet channels of the two-band 
arising from Dirac fermions in presence of the short range Coulomb interactions 
are given by,
\be
   1\pm U \chi^0(\bq,\omega)=0,
   \label{condition.eqn}
\ee
where the $-$ ($+$) sign corresponds to triplet (singlet) 
channel~\cite{BickersScalapino}, and the sign difference 
can be traced back to fermion anti-commutation relation. 
Note that although in the above expression, $\chi^0$ is the 
particle-hole (polarization) bubble, the (approximate)
bosonic operators which satisfy the above equation are of a 
peculiar form, which {\em does not} necessarily coincide with the precise 
particle-hole fluctuation form~\cite{JafariBaskaranJPCM}.
The non-interacting susceptibility $\chi^0$ employed in the 
above equation, is defined by,
\be
   \chi^0(\bq,\omega)=\frac{1}{N}\sum_{\bk}\frac{\bar n_{\bk+\bq}^+-\bar n_{\bk}^-}
   {\omega+i\eta-(\varepsilon_{\bk+\bq}^+-\varepsilon_{\bk}^-)}.
\ee
For the dispersion relation~\eqref{massive.eqn}, the susceptibility has been calculated 
by many authors, the analytic form of which reads~\cite{Pyatkovskiy,Kotov}:
\bearr
   &&\chi^{(0)}(\bq,\omega)=-\frac{|\bq|^2}{\pi\hbar(v_F^2|\bq|^2-\omega^2)}\left\{
   \frac{2\Delta}{\hbar}+\right. \nn \\
   &&\left. \frac{v_F^2|\bq|^2-\omega^2-4\Delta^2/\hbar^2}{\sqrt{v_F^2|\bq|^2-\omega^2}}  
   \arctan \frac{\sqrt{v_F^2|\bq|^2-\omega^2}}{2\Delta/\hbar}
   \right\}.
\eearr

%figure 2
\begin{figure}[tb]
\begin{center}
\includegraphics[width=8cm,angle=0]{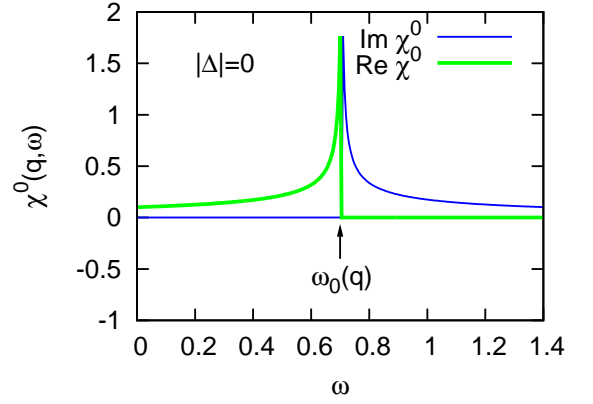}
\caption{ (Color online) Behavior of free particle hole 
propagator $\chi^0(\bq,\omega)$ when the gap vanishes, i.e. $\Delta=0$. 
The imaginary part (blue line) marks the onset of particle-hole continuum of 
free particle-hole excitations. In this figure, vertical scales are drawn 
for $qa=0.7$, and $v_F$ and $\hbar$ are assumed to be unit. 
}
\label{chi-gapless.fig}
\end{center}
\end{figure}

To understand the structure of this function, 
in Fig.~\ref{chi0.fig} we have plotted the real and imaginary part of
this function versus $\omega$ for some representative value of $\bq$.
As can be seen, the imaginary part is non-zero for $\omega > \omega_\Delta(\bq)$,
where 
\be
   \omega_\Delta(\bq)=\sqrt{v_F^2|\bq|^2+4\Delta^2/\hbar^2},
\ee
corresponds to the lower boundary of PHC of massive Dirac fermions.
The singularity of the above inter-band response at $\omega=\omega_\Delta(\bq)$
is of a weak logarithmic form which could give rise to a solution
for Eq.~\eqref{condition.eqn} only in the triplet channel, and for 
extremely large values of $U$ which is unphysical. The $\chi^0$ however 
possesses another much more interesting simple pole structure at a 
lower energy scale, $\omega=\omega_0(\bq)$ given by,
\be
   \omega_0(\bq)=v_F|\bq|.
\ee
Note that this equation defines a line in the plane of $\omega$ and $|\bq|$,
which corresponds to the lower boundary of the PHC of massless ($\Delta=0$)
Dirac fermions and is determined by the characteristic velocity, $v_F$ of the
underlying Dirac fermions. This could be considered as a property of a "parent" 
massless Dirac system which manifests itself as a simple-pole structure
when a mass parameter, $\Delta$ is turned on in the non-interacting 
part of the Hamiltonian. At the gapless point, where
$\Delta=0$, the two energy scales at $\omega_0(\bq)$ and $\omega_\Delta(\bq)$
merge and the resulting singularity of $\chi^0$ will be of inverse-square-root
form~\cite{BaskaranJafari}. 

As can be seen, to search for solutions of Eq.~\eqref{condition.eqn},
one has to look for the intersection of the real part (green line) with
the constant horizontal line $\mp\frac{1}{U}$, where the upper (lower) sign corresponds
to singlet (triplet) channel. As can be very clearly seen in the 
figure, near $\omega_0(\bq)=v_F|\bq|$, where the imaginary part of $\chi^0$
is zero, there can be poles both in singlet and triplet channel. 
The simple pole structure of the particle-hole propagator near this
energy scale implies that the poles exist for every value of
the Hubbard $U$, no matter how small or how large it is. To  
see this more clearly, note that the real part 
of the susceptibility near $\omega_0(\bq)$  behaves as,
\be
   \mbox{Re}\chi^0(\bq,\omega)\approx \frac{|\bq|\Delta}{\pi v_F\hbar^2 \left[\omega-\omega_0(\bq)\right]}.
\ee
The solutions of the eigenvalue equations~\eqref{condition.eqn} 
for arbitrary value of $U$ define two collective branches
in singlet and triplet channel as,
\be
   \omega_{\mbox{singlet/triplet}}(\bq)=v_F q
   \left( 1\mp \frac{U\Delta a^2}{\pi\hbar^2 v_F^2} \right),
   \label{poles.eqn}
\ee
where upper sign corresponds to the singlet channel, and the
lower sign corresponds to the triplet channel.
These solutions are available even for arbitrarily small of $U$. 
This formula can be interpreted as the spin charge 
separation in the sense that the singlet and the spin-1 modes
move at different velocities given by, 
\be
   v_{c/s}=v_F\left( 1\mp \frac{U\Delta a^2}{\pi\hbar^2 v_F^2} \right)
   \label{velocities.eqn}.
\ee
{\em The mere existence and simple pole structure of the $\chi^0$ 
at the energy scale $\omega_0$ is a unique consequence of the 
gap opening in the single-particle spectrum of excitations}. 
The presence of
gap in the single particle spectrum, pushes the continuum of free
particle-hole pairs to higher energies, and hence the above collective
excitations which appear around $\omega_0(\bq)=v_F|\bq|$ line are well 
separated from the boundary of PHC and therefore will be protected from Landau
damping to the incoherent background of free particle-hole excitations. 
%Note that the line $\omega_0(\bq)=v_F|\bq|$ corresponding to the lower 
%boundary of the PHC of massless Dirac fermions is indeed characterized
%by the Fermi velocity $v_F$ of the underlying Dirac fermions.

\section{Discussion}
The opening of a single particle gap in the spectrum of excitations
in graphene, although pushes the lower edge of the particle-hole continuum
from the energy scale $\omega_0$ to energy scale $\omega_\Delta$, but
still leaves behind a signature of underlying massless Dirac fermions
in the form of a singular behavior for $\chi^0$ on the line defined by 
$\omega_0(\bq)=v_F|\bq|$ corresponding to the PHC boundary of the underlying 
massless Dirac fermions.
The simple pole left on this line has two consequences: (i) Due to 
the sign change of the susceptibility across $\omega_0$, there will
be collective mode solutions at both singlet and triplet channels, 
for arbitrary values of the Hubbard $U$. The explicit form of the
solutions are given by Eq.~\eqref{poles.eqn}.
(ii) The different velocities for the two modes is reminiscent
of the situation one encounters in one-dimension, where a faster
"spin mode" overtakes the slower "charge mode". 
Albeit the difference
is that in 1D the divergences in particle-hole bubbles are of the 
characteristic inverse square root type. In the case of 2D massive Dirac fermions, 
where $\Delta$ is finite, the divergence in the particle-hole propagator is of a 
simple pole form. When the limit $\Delta\to 0$ is taken, the simple pole 
merges with the logarithmic singularity at the boundary of PHC, $\omega_\Delta$, 
and gives rise again to a inverse-square-root behavior for gapless Dirac fermions 
in 2D~\cite{BaskaranJafari}. It is therefore the non-zero value of 
$\Delta$ which provides a solution in the singlet channel in addition
to the known solution of the triplet channel at $\Delta=0$ 
point~\cite{BaskaranJafari,JafariBaskaranJPCM}.
In this sense, it appears that
the non-zero value of $\Delta$ enhances the separation of the triplet and
the singlet modes. Therefore the smallest value of the gap parameter,
$\Delta$, totally changes the nature of collective excitations in graphene.
This observation may have implications for novel approaches to the 
bosonization of the 2+1 dimensional massive Dirac fermions. 
Note that if the interaction employed was of a long-range Coulomb form,
the solution in the spin-1 channel would be of a linearly dispersing
gapped form, and the solution in the singlet channel would not exist at all.
Therefore the two modes discussed here are peculiar feature of short
range interactions, which due to remarkable value of the Hubbard $U$ 
in graphene, may have relevance to physically fabricated samples of
gapped graphene.

Now let us discuss the nature of the phase transition marked 
by vanishing of the velocity of the singlet mode. As can be seen
in Eq.~\eqref{velocities.eqn}, when the Hubbard $U$ is large-enough
to satisfy,
\be
   U_c \Delta = \frac{3\pi}{4} t^2,
   \label{Uc.eqn}
\ee
the velocity of singlet excitations becomes zero, and the
low-energy sector is exhausted by only triplet excitations of the form
\be
   \omega_{\rm triplet}(\bq)=2v_F q.
   \label{2vF.eqn}
\ee
Thinking in the spirit of slave-boson approach~\cite{Vaezi},
an electron (hole) can be assumed to be composed of its spin part, the spinon,
and the charge part the doublon (holon). Therefore a composite
object constructed from an electron and a hole can be thought of
a separate spin-1 composite of two spinons, and a charge composite
of a doublon and a holon giving rise to a spin zero boson. 
Hence the triplet mode can be interpreted as a triplet
bound state of two spinons~\cite{BaskaranJafari,JafariBaskaranJPCM},
and the new singlet mode the emergence of which is being facilitated
by a non-zero gap in the single-particle spectrum, can be interpreted as a composite
boson constructed from a doublon and a holon the total charge of which
is zero, as it should be. In this sense, the vanishing of the
velocity of singlet mode can be associated with infinite enhancement
of the effective mass of charge bosons. This gives rise to localization
of charge carriers and hence is a Mott transition~\cite{Jafari2009}. Indeed the 
decreasing trend $U_c\propto \Delta^{-1}$ as a function of the gap parameter 
$\Delta$ in Eq.~\eqref{Uc.eqn} is in qualitative agreement with 
our previous DMFT result for the phase boundary of the Mott insulating 
phase in the ionic Hubbard model~\cite{moradDMFT}. Note that this 
agreement holds for very small to moderate values of $\Delta$.
For large values of $\Delta$, the phase boundary in DMFT will be given
by $U \propto \Delta$~\cite{moradDMFT}. However, as long as experimental realization
of gap parameter in graphene is concerned, $\Delta$ will be on the
scale of few tens of meV, which is quite small in the scale of 
the hopping term $t\sim 2.8$ eV.
Moreover, the fact that the only low-energy excitations left 
in the system when $U$ goes beyond $U_c$ is a another evidence
that at $U_c$ the systems becomes a Mott insulator. Because in the
Mott insulator, the only possible low-energy modes are spin
excitations.

When $U$ is further increased, it can be seen in Fig.~\ref{chi0.fig} that 
the real part of $\chi^0$ and $+\frac{1}{U}$ (triplet channel) will intersect 
in another point which is immediately below the edge of the PHC, 
$\omega_\Delta(\bq)$, and therefore a second triplet mode at 
slightly higher energy than the first one is expected to form,
which is anticipated due to weak logarithmic singularity at $\omega_\Delta$. 
However, this mode is not likely 
to be realized as it requires quite a large value of $U$ which falls 
already in the Mott insulating phase, where the ground state maybe 
totally deformed with respect to the initial starting point of 
massive Dirac fermions employed in this work.

It is interesting to note that, although the velocity of the 
underlying Dirac fermions is $v_F$, the velocity of the spin-1
mode is always more than $v_F$. Specially at the transition to Mott
insulating phase, the velocity of the spin-1 mode will be
$2v_F$, i.e. the triplet mode (if we call it {\em triplon}) 
moves twice faster than the underlying Dirac electrons. This prediction 
can be tested not only in gapped graphene system, but also 
on a platform based on cold atoms engineered 
to mimic a massive Dirac theory at nano-Kelvin energy scales.
The existence of a mode whose velocity can be twice the velocity
of underlying Dirac fermions may have interesting implications
in teleportation of quantum information, as well as in the
spin-only forms of transport. 

The above two modes discussed here exist for small values 
of $U$ too. Therefore it should be possible to check for their
experimental consequences. The well isolation of the energy scale
$\omega_0$ from the PHC edge $\omega_\Delta$ makes the gapped
graphene more interesting for performing neutron scattering 
experiments. In this work we find that the triplet excitation
exists below the Mott transition. Moreover, the Mott insulating
phase has its own magnetic excitations. Therefore in the gapped
graphene we anticipate a triplet excitation over a large range 
of values of the Hubbard parameter $U$. Such low-energy spin-1 
branch of excitations will have a characteristic $T/v_s^2$ contribution
in the specific heat at constant volume, where $v_s$ is the velocity of spin
modes. Below the Mott transition, there will be another $T/v_c^2$ contribution
coming from singlet charge modes. 
The influence of these modes on various properties of
gapped graphene remains to be investigated.

\section{acknowledgments}
This work was supported by the National Elite Foundation (NEF) of Iran.
We are grateful to K. Yamada for hospitality at the Institute for Materials
Research, Tohoku University, and the suggestion of the present calculation.
We thank S. Rouhani and A. Vaezi for useful discussions.

\end{document}